# Effect of exclusion criteria on the distribution of blood test values

Rina Kagawa[1], Masanori Shiro[2]

[1]University of Tsukuba, Japan
[2]Advanced industrial science and technology, Japan

Corresponding author: Masanori Shiro shiro@aist.go.jp (Tel: 81-29-861-4189)

**Abstract:** The increasing demand for personalized health care has led to the expectation that individualized quantitative evaluation of human disease states is possible. However, this has not yet been achieved at a sufficiently low cost. Our ultimate goal is to determine the most accurate distributions of blood tests commonly used in health checkups. In this study, we quantified differences between the estimated distributions based on four datasets using the lognormal distribution with three parameters and analyzed the cause of the differences. We focused on two causes of differences: the exclusion criteria and distribution used for estimation of distributions. We compared the expected values across datasets for each laboratory test. We also quantitatively evaluated differences in the shape of the estimated distribution corresponding to the exclusion criteria. We found that exclusion criteria have an important influence on the shape of the distribution for blood test values.

## 1 INTRODUCTION

### 1.1 Background

Increasing demand for personalized health care, accumulation of medical data with the spread of electronic health records worldwide, and many data-intensive genomic studies in precision medicine have led to the expectation that individualized quantitative evaluation of human disease states is possible [1]. However, this has not yet been achieved at a sufficiently low cost [2]. One hurdle arises from the observed data itself. For example, in clinical practice, vital sign measurements and laboratory tests are conducted at irregular intervals [3], causing difficulties in developing a methodology for evaluating clinical status based on these data [4].

Although reference values for comparison are necessary for quantitative evaluation of observed values, it has not been fully established how the reference values should be determined. Our goal is to establish reference values for comparisons with observed values. We hypothesize that the expected value of the distribution obtained from the healthy population is the best reference for comparison because the expected value could represent the healthiest condition and deviations from the expected value can be calculated and used to evaluate changes in patient's health status.

Our study focuses on laboratory tests for evaluating disease status. A reference for comparison must be set for each laboratory test. Laboratory test measurements must be evaluated for deviations, including mean absolute deviation and standard deviation from the reference. However, distributions obtained from the healthy population have not been published for most laboratory tests; therefore, we need to assume these distributions. For calculation of deviations at low cost, a parametric distribution is desirable because the expected and integral values of parametric distributions can be calculated quickly with high accuracy if the parameters of the distribution can be estimated accurately.

The objective of this study is to clarify the parametric distribution and parameters of the reference distribution for laboratory tests. However, the two endpoints of the reference interval (RI) and the median (referred to as *the three values*) have been published for only some laboratory tests [5][6]. For a few laboratory tests, histograms have been published as well [7], but the expected value cannot be calculated accurately from the histogram because the widths of the histogram's bins were arbitrarily determined. Moreover, these histograms and previous studies have demonstrated that many laboratory test values from multiple hospitals have asymmetric distributions [8]. Since the expected value and the median are different in an asymmetric distribution, the expected value of the distribution is difficult to calculate based only on the three values. Therefore, deviations of an observed value from the asymmetric distribution cannot be accurately calculated.

### 1.2 Clarification of the results of a previous study and the purpose of this study

In our previous study [9], we selected a parametric statistical distribution that represented the distribution of a laboratory test value with the smallest error using an actual hospital dataset. We examined 11 typical unimodal continuous distributions, which are provided as standard distributions in statistical software R or are modified distributions provided in R by adding the degree of freedom along the horizontal axis. These distributions include the normal distribution (norm), lognormal distribution, Cauchy distribution, chi-square distribution, F distribution, gamma distribution, logistic distribution, Poisson distribution, Weibull distribution, and three-parameter lognormal distribution (lnorm3). An inverse conversion of the modified Box-Cox transformation used by Ichihara and Japanese Committee for Clinical Laboratory Standards (JCCLS) was also examined. These 11 statistical distributions have two or three parameters each.

Many laboratory test values follow the lnorm3 distribution, which was consistent with results from previous studies [10] that showed the lognormal distribution can be used to estimate the distribution of test values. The density function of the lnorm3 distribution is shown in Eq.1.

$$f(x; \mu, \sigma, d) = \frac{1}{\sqrt{2\pi\sigma^2}} \exp\left(-\frac{(x-d-\mu)^2}{2\sigma^2}\right) \quad \text{(Eq. 1)}$$

$$f(x; \mu, \sigma, d) = \frac{1}{\sqrt{2\pi\sigma^2}(x-d)} \exp\left(-\frac{(\log(x-d)-\mu)^2}{2\sigma^2}\right) \quad \text{(Eq. 2)}$$

To further validate our previous study [9], we compared normal distributions estimated using the three parameters (norm3) and lnorm3, each with three degrees of freedom. Some results are shown in Figure 1. The black dots represent kernel density estimate (KDE), the green line represents the distribution estimated using norm3, and the red line represents the distribution estimated using lnorm3. The density function for norm3 is shown in Eq. 2. For all blood tests in our previous study [9], the distribution estimated using lnorm3 was equivalent or had a better fit to the KDE than norm3.

Our previous study confirmed differences between the lognormal distributions estimated based on data collected in two actual hospitals or health screening projects in Japan versus those based on previous research. Previous studies have pointed out that the RIs of a distribution differ by gender [5][6][10]. However, from our previous results, we inferred that differences in properties of the four datasets also affect the shape of the distributions as much as or more than gender.

Therefore, in this study, we quantified the differences between the estimated distributions based on the datasets and discussed the cause for the differences.

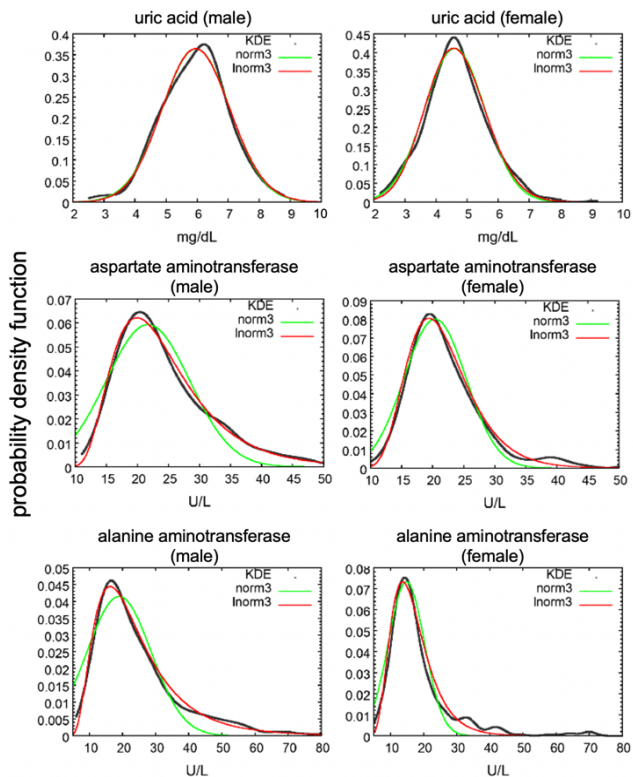

*Figure 1: Examples of distributions estimated using norm3 and lnorm3 compared with KDE.*

## 2 Materials and Methods for distribution estimation

**Materials**

This study used four datasets, similar to our previous study [9]. The laboratory tests included in each dataset are shown in Table 1. The abbreviations for laboratory tests are drawn from a previous study [11]. Results were stratified by gender for each laboratory test. Unless otherwise stated, exclusion criteria were not defined for these datasets.

### 1. University of Tsukuba Hospital (UTH)

Laboratory test values during medical checkups were obtained for 518 males (M) and 512 females (F) over the age of 20 years (average age, 60.72 years) between January 1, 2017 and December 31, 2018 at UTH. For persons who had multiple medical check-ups, the value from the first check-up for each laboratory test was used for the analysis.

### 2. National Health and Nutrition Survey (NHNS)

NHNS is a health screening project in Japan. The histograms of some laboratory tests have been published [7]. We used data from 2013 to 2019, except for 2016 because most data were missing according to the data source policy [7]. Values from 7,632 males and 10,796 females over the age of 20 years were used. The values of the histogram boundaries for each laboratory test were the same for every year. For easy comparison with other datasets, we obtained one histogram by averaging the frequencies contained in each bin of the original annual histogram, which was normalized for each laboratory test item.

### 3. Japanese Committee for Clinical Laboratory Standards (JCCLS)

This dataset contains three laboratory values of Japanese individuals (2,733 males and 3,612 females, ages 18–72 years) [6][11]. Exclusion criteria were based on parameters such as body mass index and alcohol intake. JCCLS was designed to recruit healthy subjects [6] because RIs are generally calculated based on a healthy population.

To create this dataset [6][11], histograms obtained via measurements were converted into normal distributions with the modified Box-Cox transformation [12]. Values corresponding to 5% and 95% intervals were obtained. RIs were then obtained via inverse transformation. The density function $h(x)$ that we used explicitly is shown in Eq. 3.

$$h(x) = \frac{(x - \mu + 4\sigma)^{p-1}}{\sqrt{2\pi}\sigma} \exp\left(-\frac{\left(\frac{(x - \mu + 4\sigma)^p - 1}{p} - m\right)^2}{2\sigma^2}\right)$$

(Eq. 3)

Due to the shift in $x$ without the expected value $\mu$ being fixed as $4\sigma$, some types of histograms might not be fitted to $h(x)$ appropriately in Eq. 2. The errors caused by these transformations would affect published RIs.

*Table 1– Laboratory tests*

| Item | Dataset | | | |
|---|---|---|---|---|
| | UTH | NAHS | JCCLS | Ichihara |
| red blood cell count (RBC) (×$10^6$/μL) | ✓ | | ✓ | |
| hemoglobin (Hb) (g/dL) | ✓ | | ✓ | |
| hematocrit (Ht) (%) | ✓ | | ✓ | |
| albumin (Alb) (g/dL) | ✓ | | ✓ | ✓ |
| total protein (TP) (g/dL) | ✓ | | ✓ | |
| aspartate aminotransferase (AST) (U/L) | ✓ | ✓ | ✓ | ✓ |
| alanine aminotransferase (ALT) (U/L) | ✓ | ✓ | ✓ | ✓ |
| gamma-glutamyl transpeptidase (g-GTP) (U/L) | ✓ | ✓ | ✓ | ✓ |
| alkaline phosphatase (ALP) (U/L)[1] | ✓ | | ✓ | |
| amylase (AMY) (U/L) | ✓ | | ✓ | ✓ |
| lactate dehydrogenase (LD) (U/L) | ✓ | | ✓ | ✓ |
| total cholesterol (Cho) (mg/dL) | ✓ | ✓ | ✓ | ✓ |
| triglycerides (TG) (mg/dL) | ✓ | ✓ | ✓ | ✓ |
| HDL-cholesterol (HDL) (mg/dL) | ✓ | ✓ | ✓ | ✓ |
| LDL-cholesterol (LDL) (mg/dL) | ✓ | ✓ | ✓ | ✓ |
| creatine kinase (CK) (U/L) | | | ✓ | ✓ |
| creatinine (Cr) (mg/dL) | ✓ | ✓ | ✓ | ✓ |
| uric acid (UA) (mg/dL) | ✓ | ✓ | ✓ | ✓ |
| sodium (Na) (mEq/L) | ✓ | | ✓ | ✓ |
| potassium (K) (mEq/L) | ✓ | | ✓ | ✓ |
| chloride (Cl) (mEq/L) | ✓ | | ✓ | ✓ |
| calcium (Ca) (mg/dL) | ✓ | | ✓ | ✓ |
| hemoglobin A1c (HbA1c) (%) | ✓ | | ✓ | |
| fasting blood sugar (Glu) (mg/dL) | ✓ | | ✓ | |
| IgA antibody (IgA) (mg/dL) | | | ✓ | ✓ |
| IgG antibody (IgG) (mg/dL) | | | ✓ | ✓ |
| IgM antibody (IgM) (mg/dL) | | | ✓ | ✓ |
| C3 (mg/dL) | | | ✓ | ✓ |
| C4 (mg/dL) | | | ✓ | ✓ |

[1] the method originally used in Japan was used.

### 4. Ichihara+ 2013 (Ichihara)

Ichihara et al. reported three values for each laboratory test in 2,082 Japanese individuals aged 20–65 years [5]. Similar to JCCLS, RIs were calculated for this dataset using the modified Box-Cox transformation. Exclusion criteria were similar to those for JCCLS.

Previous work, in which the median of each laboratory test has not been published [13], were excluded from our experiments. A national medical study conducted in Japan [14] was also excluded because the published data were biased and distribution parameters could not be estimated.

### Methods for estimation of distributions

Prior to the experiments, we used lnorm3 to estimate the CUTH[1], JCCLS, and Ichihara distribution parameter sets for each of the 58 laboratory tests stratified by gender. For CUTH and NAHS, the sum of the squared errors for each laboratory test was calculated between the value at the center of each column in the histogram and a point at the same location in the estimated distribution. In other words, the parameter set of the distribution was determined using the Nelder-Mead method to minimize the sum of the squared errors. For JCCLS and Ichihara, the three values were provided for each laboratory test, allowing for unique determination of the parameter set for lnorm3.

Two experiments were performed based on the estimated parameter sets for each laboratory test in each dataset.

The following computing environments were used: R 3.6.3, Perl 5.30, Python 3.7.2, and SciPy 0.6.6 with Ubuntu 20.04 LTS.

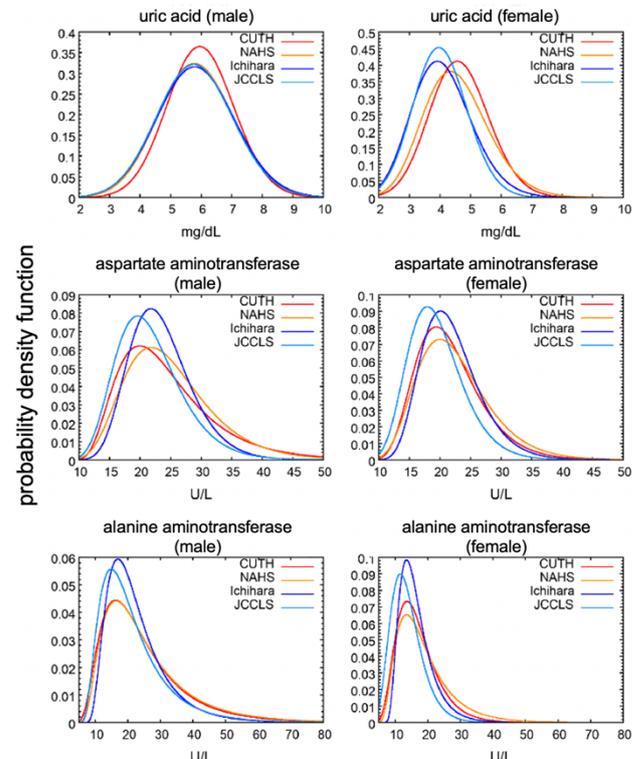

*Figure 2: Examples of estimated distributions using lnorm3 for the four data sets.*

converted using KDE.

[1] We used the label *CUTH* for the UTH distribution

## 3 Experiment 1: Comparison of expected values across datasets for each laboratory test

Our previous work [9] showed that CUTH and NAHS had similar estimated distributions, and Ichihara and JCCLS had similar estimated distributions, but these two groups of estimated distributions were not similar to each other.

To clarify the cause of differences in estimated distributions across datasets, we focused on two features common to Ichihara and JCCLS: (1) similar exclusion criteria and (2) use of the modified Box-Cox transformation. For Ichihara and JCCLS, RIs were estimated using the inverse transformation of the modified Box-Cox transformation after converting the original measurements to a normal distribution [5][6]. Ichihara and JCCLS were based on common exclusion criteria, used to recruit healthy subjects. These exclusion criteria have not been applied to CUTH. If this has a dominant influence on the estimation of distributions, the shape of the estimated distributions for Ichihara and JCCLS would be interpreted as healthier than the shape of estimated distributions for CUTH. For example, the peak for AST (M) with Ichihara would be smaller than that with CUTH.

In this study, we investigate the possibility that exclusion criteria are the cause of differences in estimated distributions across datasets.

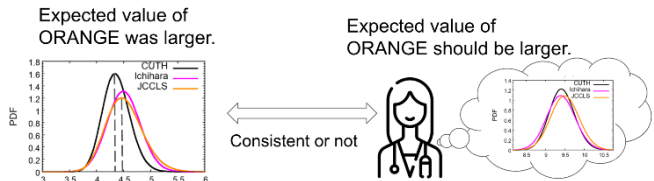

*Figure 3: Overview of the comparison of expected values between observed values and physician knowledge (Experiment 1).*

### 3.1 Methods

To test the hypothesis that different exclusion criteria among datasets are the cause of differences in estimated distributions across datasets, we focused on the expected value of the distribution. If exclusion criteria are the main cause of differences in distributions, then the direction of difference in expected values of CUTH versus JCCLS and Ichihara would be consistent with physicians' expectations based on their medical knowledge of the populations in each dataset over many laboratory tests. We determined whether the direction of the difference in the expected value (Eq. 4) of the lnorm3 distribution for Ichihara or JCCLS is larger or smaller than the expected value for CUTH. We call this direction of difference from the expected value "DFE."

$$\exp\left(\mu + \left(\frac{\sigma^2}{2}\right)\right) + d \quad \text{(Eq. 4)}$$

Moreover, one of the authors, a physician, predicted DFEs based on common clinical knowledge and experiments. The DFE calculated based on our estimated distribution and the physician-predicted DFE was compared for each laboratory test.

### 3.2 Results

Among the 46 laboratory tests by gender, DFE calculated based on our estimated distributions and physician-predicted DFE agreed for all except the following four tests by gender: AST (F), Hb (F), Ht (F), and ALP (M) (Table 2). The direction of the difference in the expected value of the estimated distribution for JCCLS and Ichihara compared to CUTH tended to be due to the exclusion criteria.

*Table 2: Laboratory tests in which DFE and physician-predicted direction of change do not match*

|  | Consistency between DFE and physician-predicted direction of change | |
|---|---|---|
| Laboratory tests | *JCCLS* | *Ichihara* |
| AST (F) | ✓ | X |
| Hb (F), Ht (F), ALP (M) | X | ✓ |

## 4 Experiment 2: Distance between the estimated distributions of two datasets

The results of Experiment 1 suggested that differences in the shapes of the estimated distributions were due to the exclusion criteria. In Experiment 2, we quantitatively evaluated differences in the shape of the estimated distribution corresponding to the exclusion criteria.

### 4.1 Methods

For any pair of estimated distributions, we defined the distance as the sum of the absolute values of the differences in the probability density functions (PDFs) of the two distributions. Specifically, we calculated the sum of the squared differences taken at 10,000 points at the same location for the distributions.

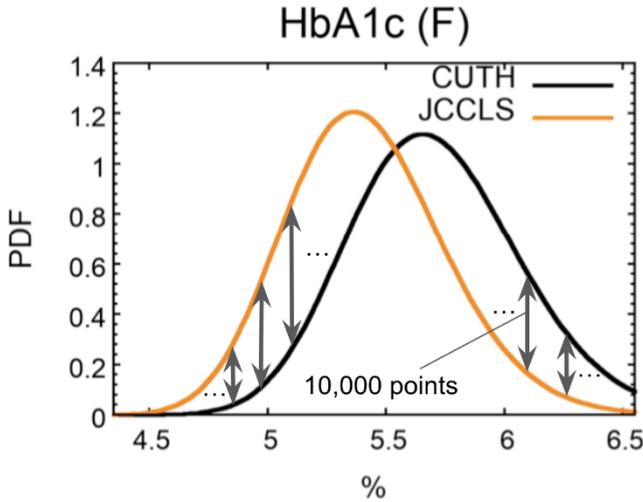

*Figure 4: Overview of the evaluation of the distance between estimated distributions between two datasets (Experiment 2). The sum of the squared difference was calculated at 10,000 points between each pair of distributions.*

## 4.2 Results

The expected value of the difference in PDF between Ichihara and JCCLS, CUTH and NAHS, Ichihara and CUTH, Ichihara and NAHS, JCCLS and CUTH, and JCCLS and NAHS was 0.17, 0.19, 0.32, 0.38, 0.32, and 0.38, respectively.

We divided these six pairs of estimated distributions into two groups based on the expected value of the difference between PDFs. Group 1 consisted of two pairs: Ichihara and JCCLS, and CUTH and NAHS. Group 2 consisted of Ichihara and CUTH, Ichihara and NAHS, JCCLS and CUTH, and JCCLS and NAHS. These two groups are composed of two pairs with similar exclusion criteria and four pairs with dissimilar exclusion criteria. We refer to Group 1 as "Near" and Group 2 as "Far." Figure 5 shows the distribution of relative distances in the Near group (average distribution of distances between Ichihara and JCCLS and between CUTH and NAH) and the Far group (average distribution of distances for the other four pairs) for each laboratory test using KDE. The horizontal axis represents distance. The vertical axis represents PDF, or the frequency of distances. The sample sizes were 62 (Near) and 114 (Far). The average PDF values were 0.335 (Far) and 0.179 (Near). The null hypothesis that the average PDF values of the Near and Far groups are equal was rejected ($p < .001$; effect size = 0.767).

Thus, the shapes of the four datasets were divided into two groups: CUTH and NAHS, and JCCLS and Ichihara. The datasets within each group were quantitatively shown to be similar to each other.

The small peak around 1.0 for Near is due to the fact that the RIs for g-GTP in Ichihara are different from those in JCCLS [10, 12]. Small peaks were observed around 0.5 for Near and Far, but this is difficult to explain. As shown in Figure 5, the relative distance for many pairs of two distributions were less than 0.8. Since the relative distance is the sum of the absolute difference for 10,000 points, the difference in the PDF per point is considered to be within $8\times10^{-5}$ on average. Therefore, the expected difference in PDF per point should be assumed to be approximately $8\times10^{-5}$ when the exclusion or inclusion criteria for population selection used to obtain the three values required to estimate the distribution with lnorm3 is unknown. The values of the two RI endpoints depend on the distribution's shape. However, it is generally not possible to estimate how two RI endpoints change based on the shape of the distribution.

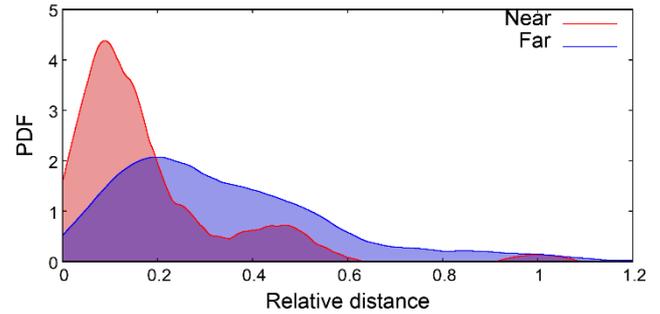

*Figure 5: Distribution of relative distance as the sum of the absolute value of differences between two estimated distributions.*

## 5 Discussion

We calculated the distance between estimated distributions using the sum of the absolute value of differences, but the distance could also be estimated using other measures such as symmetrical Kullback-Leibler divergence. We showed that the use of statistical distributions other than the lognormal distribution should not be precluded. We also showed that even when the original data were examined by gender, the influence of the criteria for selecting the population, which affects the shape of the distribution of laboratory test values, cannot be ignored. Therefore, the distribution should be estimated based on well-defined inclusion or exclusion criteria. However, a small sample size has a non-negligible effect when determining the shape of a distribution. One way to solve this dilemma is to agree on a well-defined criterion. However, it is difficult to reach a broad consensus, even for a specific purpose, such as recruiting healthy subjects. Data accumulation is progressing rapidly, and a distribution can be determined stably by taking a large sample size without defining exclusion or inclusion criteria.

**Limitations and future work**

The data used in this study were acquired exclusively in Japan. Another limitation is that the four datasets did not include individuals in the same age range. For one laboratory test, when the median value of the distribution is unknown and only RIs have been published, our method would not be applicable. Even if the three values of the distribution of a laboratory test are given, when two or three of them are similar, it is not possible to determine the three parameters of the distribution with much clarity [15]. A future study could determine what types of two-parameter distributions are appropriate for such laboratory tests. Another future study could further clarify the effect of the population selection criteria on distributions. For example, we will try to estimate the distribution obtained from inpatients and provide detailed analysis, including stratification by disease. In other future work, we will simultaneously estimate the shapes of distributions of pairs of laboratory tests.

# 6 Conclusion

We clarified that differences in the shapes of estimated distributions are due to exclusion criteria. We also clarified differences between estimated distributions based on three values when no criteria have been published. The results of this research are expected to be incorporated into several clinical tests. We believe that the results will be the foundation for constructing reference values for quantitative evaluation of human disease states in the future.

## Ethics

This research was approved by the ethics committee of the University of Tsukuba Hospital (approval number: R1-080).

## Acknowledgements

The present study was supported in part by the Japan Science and Technology Agency (JST)-Mirai Program Grant Number JPMJMI19G8, the Japan Society for the Promotion of Science (JSPS), Grants-in-Aid for Scientific Research (Nos. JP18H06363, JP19K19347), R&D Center for Frontiers of MIRAI in Policy and Technology, Healthcare-Medical Fusion Project in AIST and TIA nano KAKEHASHI Tsukuba Innovation Grants. The funders had no role in the study design, data collection and analysis, decision to publish, or preparation of the manuscript. Ms. Emiko Nishida, Ms. Mika Sumimoto, and Ms. Noriko Ohkubo have greatly contributed to manual data processing. The free icon figure is from flaticon.com.